\documentclass[11pt]{article}

\usepackage{euscript}
\usepackage{amssymb}
\usepackage{amsfonts}
\usepackage{amsbsy}
\usepackage{amsmath}
\usepackage{epsfig}
\usepackage{amsthm}
\usepackage{amscd}
\usepackage{amstext}
\usepackage{graphicx,dsfont,multirow,ctable,colortbl,rotating}

\usepackage[pdftex]{hyperref}
\makeatletter
\renewcommand\section{\@startsection {section}{1}{\z@}%
                                   {-3.5ex \@plus -1ex \@minus -.2ex}
                                   {2.3ex \@plus.2ex}%
                                   {\normalfont\large\bfseries}}

\renewcommand\subsection{\@startsection{subsection}{2}{\z@}%
                                     {-3.25ex\@plus -1ex \@minus -.2ex}%
                                     {1.5ex \@plus .2ex}%
                                     {\normalfont\bfseries}}

\newlength{\apb@width}
\newcommand{\autoparbox}[2][c]{\settowidth{\apb@width}{#2}\parbox[#1]{\apb@width}{#2}}

\makeatother

\newcommand{\bea}{\begin{eqnarray}}
\newcommand{\eea}{\end{eqnarray}}
\newcommand{\be}{\begin{equation}}
\newcommand{\ee}{\end{equation}}

\newcommand{\bem}{\begin{pmatrix}}
\newcommand{\eem}{\end{pmatrix}}

\def\a{\alpha}

\def\c{\gamma}
\def\d{\delta}

\def\e{\epsilon}   
\def\f{\phi}               

\def\inf{\infty}

\def\l{\lambda}

\def\m{\mu}
\def\n{\nu}

\def\p{\pi}   
\def\pa{\partial}       
\def\r{\rho}                                     
\def\s{\sigma}                                   
\def\t{\tau}
\def\th{\theta}
\def\til{\tilde}

\def\D{\Delta}
\def\F{\Phi}
\def\G{\Gamma}

\def\L{\Lambda}
\def\O{\Omega}

\def\Tr{{\rm Tr}}

\def \Z {{\mathbb Z}}

\def \R {{\mathbb R}}

\begin{document}
\begin{center}

\vspace{1cm} {     \LARGE {\bf  K3 Surfaces, ${\cal N} =4 $ Dyons, \\ \vspace{0.2cm}and   the Mathieu Group $M_{24}$}}

\vspace{2cm}

Miranda C. N. Cheng
\vspace{0.7cm}

{\it Department of Physics, Harvard University,
\\
Cambridge, MA 02138, USA \\
}

\vspace{1.5cm}

\end{center}

\begin{abstract}

A close relationship between 
$K3$ surfaces and the Mathieu groups has been established
in the last century. 
Furthermore, it has been observed recently that the elliptic genus of $K3$ has a natural interpretation in terms of the dimensions of representations of the largest Mathieu group $M_{24}$. 
In this paper we first give further evidence for this possibility by studying the elliptic genus of $K3$ surfaces twisted by some simple symplectic automorphisms. These partition functions with insertions of elements of $M_{24}$ (the McKay-Thompson series) give further information about the relevant representation. 
We then  point out that this new ``moonshine" for the largest Mathieu group is connected to an earlier observation on a moonshine of $M_{24}$ through the $1/4$-BPS spectrum of $K3\times T^2$-compactified type II string theory. 
This insight on the symmetry of the theory sheds new light on the generalised Kac-Moody algebra structure appearing in the spectrum, and leads to predictions for new elliptic genera of $K3$, perturbative spectrum of the toroidally compactified heterotic string, and the index for the $1/4$-BPS dyons in the $d=4$, ${\cal N}=4$ string theory, twisted by elements of the group of stringy $K3$ isometries.

\end{abstract}

\pagebreak
\setcounter{page}{1}

\section{Introduction and Summary}
\label{Introduction and Summary}

Recently there have been two new observations relating K3 surfaces and the largest Mathieu group $M_{24}$. 
They seem to suggest that the sporadic group $M_{24}$ naturally acts on the spectrum of $K3$-compactified string theory. In this paper we will further explore this relation by studying various spectra, perturbative and non-perturbative, of $K3$-compactified string theories. 

Throughout the development of string theory, $K3$ surfaces have been playing a prominent role. Being the unique Calabi-Yau two-fold, $K3$ compactifications serve as an important playground for supersymmetric compactifications. It also features in one of the most important dualities in string theory that links heterotic strings to the rest of the string family. 

It is well-known that there are 26 sporadic groups in the classification of finite simple groups. 
The first ones discovered are the five so-called Mathieu groups. The largest one in the family is $M_{24}$, which can be thought of as acting as permutations of 24 objects. It has an index $24$ subgroup $M_{23}$, which can be thought of as acting as permutations of 24 objects with one of them held fixed. 

The relation between sporadic groups and modular objects has been a fascinating topic of mathematical research in the past few decades, connecting different fields such as automorphic forms, Lie algebras, hyperbolic lattices, and number theory. The most famous example is the so-called Monstrous Moonshine, which relates the largest sporadic (the Monster) group to modular forms of discrete subgroups of $PSL(2,\R)$. On the physics side, this topic has been deeply connected to string theory from the start. For example, the ``monstrous module" $V^\natural$ (an infinite-dimensional graded representation of the Monster group) and Borcherds' Monster Lie algebra naturally arise from compactifying bosonic string theory on an $\Z_2$-orbifolded Leech lattice. See Ref. \cite{gannon},  for example,  for a review of the topic. A possible relation with three-dimensional gravity has also been suggested recently \cite{Witten2007,Duncan2009}. In this paper we would like to explore various aspects of a possible similar ``moonshine" for the sporadic group $M_{24}$, which appears to arise from string theory probing $K3$ surfaces.

In the last century, S. Mukai \cite{Mukai} considered the finite groups which act on $K3$ surfaces and fix the holomorphic $2$-form, a condition tied to the physical condition of  supersymmetry and defining the so-called symplectic automorphisms, and showed that all of them can be embedded inside the sporadic group $M_{23}$. This relation between the classical geometry of $K3$ and the Mathieu group was later further explained by S. Kondo~\cite{Kondo} from a lattice theoretic view point.

Before discussing the recent observations relating $K3$ surfaces and the largest Mathieu group $M_{24}$, let us first comment on the possible origin of the appearance of $M_{24}$ in the elliptic genus. Since it was shown that all symplectic automorphisms of $K3$ surfaces can be embedded in $M_{23}$, the idea that the full $M_{24}$ acts on the spectrum of CFT on $K3$ seems simply excluded. The fallacy of such an argument lies in the fact that the moduli space of string theory on $K3$ surfaces includes more than just the classical geometry part. The B-field degrees of freedom for example, should definitely be taken into account. To be more specific, the analysis of Kondo~\cite{Kondo} shows that  the action of symplectic automorphisms on the lattice $H^2(K3,\Z)\simeq \G^{3,19}$ of classical $K3$ geometry can be identified with the action of certain elements of $M_{23}$ on Niemeier lattices ({\it{i.e.}} even, self-dual, negative-definite lattices of rank 24). But once the two-form field is taken into account, the relevant lattice for the CFT is in fact the larger quantum lattice $H^{2\ast}(K3,\Z)\simeq \G^{4,20}$.

Moreover, there does not have to be a point in the moduli space where $g$ generates an symmetry of the CFT for every element $g$ of $M_{24}$ in order for the elliptic genus of $K3$ to be a representation of $M_{24}$.
This is because the elliptic genus is a moduli-independent object, which implies when different symmetries are enhanced at different points in the moduli space, the elliptic genus has to be invariant under the whole group these different symmetries generate. But this does not imply that all subgroups of the group in question are realised as actual symmetries of the CFT\footnote{I thank Ashoke Sen for useful discussions on this point.}.
 
Finally, as we will see in details later, the Fourier coefficients of the $K3$ elliptic genus also have the interpretation as the root multiplicity of a generalised Kac-Moody algebra generating the $1/4$-BPS spectrum of $K3\times T^2$-compactified string theory, or equivalently $T^6$-compactified heterotic string theory. It is hence possible that the whole $M_{24}$ should better be understood in this context, where the relevant charge lattice is the larger $\G^{6,22}\oplus\G^{6,22}$.

Now we are ready to discuss the recent observations relating $K3$ and $M_{24}$. In Ref. \cite{Eguchi2010}, Eguchi, Ooguri and Tachikawa studied the following decomposition of the elliptic genus of $K3$ in the characters of representations of an ${\cal N}=4$ superconformal algebra
\bea\notag
{\cal Z}_{K3}(\t,z) &=& \Tr_{\text{RR}}\left((-1)^{J_0+\bar J_0} q^{L_0} \bar q^{\bar L_0} e^{2\p i z J_0} \right)\\\notag
&=& \frac{\th_1^2(\t,z)}{\eta^3(\t)}\,\left(24 \,\m(\t,z) +q^{-1/8}\left(- 2  + T(\t)\right) \right)\;.
\eea
They then observe that the first few Fourier coefficients of the $q$-series $T(\t)$ are given by simple combinations of dimensions of irreducible representations of the sporadic group $M_{24}$. In other words, it appears as if there is an infinite-dimensional representation, which we will call $K^\natural$, of the largest Mathieu group, that has a grading related to the $L_0$-eigenvalues of the CFT, such that 
$$
K^\natural  =\bigoplus\limits_{n=1}^\inf K^\natural_n\quad,\quad T(\t) = \sum_{n=1}^\inf q^n \, \text{dim}(K^\natural_n)
\;.
$$

It would constitute very strong further evidence for the validity of the above assumption if one can show that the corresponding McKay-Thompson series, or twisted partition functions
 $$
 T_g(\t)= \sum_{n=1}^\inf (\Tr_{K^\natural_n}g)\; q^n\quad,\quad g\in M_{24}\;,
 $$
as well play a role in the $K3$ CFT.  
This motivates us to look at the $K3$ elliptic genera twisted by generators of the simplest groups $\Z_{p=2,3,5,7}$
of symplectic automorphism groups of classical $K3$ geometry, which are all the cyclic $\Z_p$ groups in Nikulin's list of $K3$ quotients with prime $p$. These objects have been computed in Ref.~\cite{DavidJHEP0606:0642006} and we found that under similar decomposition they are indeed given by the McKay-Thompson series of the appropriate elements $g\in M_{24}$. Apart from further supporting the existence of such a $M_{24}$ module, the twisted elliptic genera also give much finer information about the representation $K^\natural$. Furthermore, using the conjectural relation between $M_{24}$ characters and twisted elliptic genera, we can now easily compute the elliptic genera twisted by various other elements of $M_{24}$. The reason for this is the following. Using standard CFT argument we expect the twisted elliptic genera to be weak Jacobi forms of congruent subgroups of $PSL(2,\Z)$. This modularity leaves just a few unknown coefficients to be fixed in the expression for the elliptic genera. The knowledge about the few lowest-lying representations $K_n^\natural$ with small $n$ together with the character table (Table \ref{M24}, Appendix \ref{Character Tables})  is sufficient to fix all unknown coefficients. The full expression can then in turn be used to determine the representation $K_n^\natural$ for higher $n$. We demonstrate this by giving explicit formulas for elliptic genera twisted by various other elements of $M_{24}$, including both those generating other geometric symplectic automorphisms and those that do not have known geometric interpretations.

Now we turn to the other recent observation regarding the perturbative spectrum of heterotic string theories. 
It is a familiar fact that the  $1/2$-BPS spectrum of the heterotic theory compactified on $T^6$ is given by nothing but the bosonic string partition function
$$
\frac{1}{\eta^{24}(\t)} = q^{-1}+\sum_{n=0}^\inf d_n q^n\quad,\quad q= e^{2\p i\t}\;,
$$
where $\eta(\t)$ denotes the Dedekind Eta function
$$
\eta(\t) = q^{1/24} \prod_{n=1}^\inf (1-q^n)\;.
$$ 

In Ref~\cite{SenAdv.Theor.Math.Phys.9:527-5582005,JatkarJHEP0604:0182006}, at points with enhanced symmetries, the partition function of the above theory twisted by an $\Z_{p=2,3,5,7}$ symmetry has been computed to be 
$$
\frac{1}{\eta^{\frac{24}{p+1}}(\t)\eta^{\frac{24}{p+1}}(p\t)}\;.
$$
Very recently, Govindarajan and Krishna \cite{Govindarajan2009}
extended the above results  to other $\Z_N$ orbifolds in Nikulin's list of $K3$ quotients. 
In their derivation, these authors have used the heterotic-type II duality and the fact that all $K3$ symplectic automorphisms can be embedded in $M_{23}\subset M_{24}$. They also observed that the resulting partition functions are given by the product of $\eta$-functions with the so-called ``cycle shape" of the corresponding elements of $M_{24}$. See Section~\ref{Review} and Table~\ref{M24} for details.

In fact, these $\eta$-products are known to be the central objects of a version of the ``moonshine" relating $M_{24}$ and cusp forms \cite{Mason}. The first observation is
$$
\eta^{24}(\t) = \sum_{n=1}^\inf \t_n q^n = \sum_{n=1}^\inf q^n \, \text{sdim}\til K_n\;,
$$
where $\t_n$ are the Ramanunjan $\t$-functions and $\til K = \oplus_{n=1}^\inf \til K_n$ is an infinite-dimensional graded representation of $M_{24}$. The Ramanunjan numbers are integers of both signs, and this is reflected in the fact that the representation should be thought of as coming in both bosonic and fermionic varieties, with positive and negative super-dimensions. For $g\in M_{24}$, the corresponding $\eta$-product is then the corresponding McKay-Thompson series
$$
\eta_g(\t)= \sum_{n=1}^\inf q^n\, \Tr_{\til  K_n} \left( (-1)^F g \right)\;,
$$
which reduces to the above formula for the Ramanunjan numbers when $g$ is taken to be the identity element. For those $g$ that generate the geometric $\Z_p$ symplectic automorphisms, the corresponding $\eta$-products are exactly those that give the above twisted heterotic string spectrum. 

A quick glimpse at the character table (Table \ref{M24}) suggests that the two $M_{24}$ modules $\til K$ and $K^\natural$ have really nothing to do with each other. Hence we have now two versions of moonshine for the largest Mathieu group, one related to the $K3$ elliptic genus and one to the heterotic string spectrum. But it turns out that they are not unrelated after all. Before explaining the physics, let us first switch gear and remind ourselves a few things about Borcherds' generalised Kac-Moody algebra. A generalised Kac-Moody algebra is an infinite-dimensional Lie algebra which has a denominator identity equating an infinite sum to an infinite product. In the example of the Monstrous Moonshine, the denominator of the Monster Lie algebra is given by the famous expression
$$
j(p)-j(q) = \left( \frac{1}{p}-\frac{1}{q}\right) \prod_{n,m=1}^\inf  \left(1- p^n q^m \right)^{d({nm})}
$$
where  the $j$-function
$$
j(q) -744= q^{-1}+196884 q +\dotsi = \sum_{n=-1}^\inf q^n \, \text{dim}(V_n^\natural)  = \sum_{n=-1}^\inf d(n)\,q^n  
$$
is related to an infinite dimensional graded representation $V^\natural=\oplus_{n=-1}^\inf V_n^\natural $ of the Monster group. 
The multiplicity of the positive root corresponding to the factor $(1-p^n q^m)$ inside the infinite product is given by the Fourier coefficients $d(nm)=\text{dim}(V_{nm}^\natural)$ of the $j$-function. One can also twist the above expression by elements of the Monster group and obtain the so-called twisted denominators which have played an important role in Borcherds' proof of the moonshine conjecture \cite{borcherds_monstrous}. More precisely, the product expression for the denominator twisted by an element $g$ of the Monster is given by the Fourier coefficients of the McKay-Thompson series $ \sum_{n=-1}^\inf q^n \, \Tr_{V_n^\natural}(g^k)$, where $k$ runs from one to the order of $g$.

It turns out that a similar structure is present in our $M_{24}$ case as well. It has been proposed that the $1/4$-BPS dyon spectrum in the ${\cal N}=4$, $d=4$ theory obtained by compactifying type II string on $K3\times T^2$ is generated by a generalised Kac-Moody superalgebra \cite{Cheng2008a,DVV}. 
The multiplicities of the roots of the algebra are given by the Fourier coefficients  of the elliptic genus of $K3$
$$
{\cal Z}_{K3}(\t,z) = \sum_{n\geq 0 , \ell\in \Z } c(4n-\ell^2)q^n y^\ell\;,
$$
and are hence given by the representation $K^\natural$. These data give the denominator of the algebra
$$
\F(\r,\s,\n) = e^{2\p i (\r+\s-\n)}
\prod_{\substack{n,m\geq 0, \ell \in \Z\\ \ell>0 \text{ when }n=m=0 }} (1- e^{2\p i (n\r+m\s+\ell\n)})^{c(4nm-\ell^2)}\;.
$$
which gives the $1/4$-BPS spectrum of the theory. For consistency, the $1/2$-BPS spectrum, in this case given by another representation $\til K$, should also be encoded in this $1/4$-BPS spectrum. This is because the wall-crossing phenomenon in the ${\cal N}=4$, $d=4$ theory dictates that the $1/4$-BPS spectrum jumps across the walls of marginal stability in the moduli space where $1/4$-BPS bound states decay into a pair of $1/2$-BPS particles. 
Indeed, at its poles 
$$\lim_{\n\to0} \frac{1}{\F(\r,\s,\n)}  = -\frac{1}{4\p^2\n^2}  \frac{1}{\eta^{24}(\r)}\frac{1}{\eta^{24}(\s)}
$$
the $1/4$-BPS partition function factorises into two $1/2$-BPS partition functions and correctly reproduces the expected jump in the indices \cite{Sen2008,Cheng2008a}. 
Hence, rather remarkably these two different versions of moonshine for $M_{24}$ both make their appearance in the denominator of the dyon algebra. 
Turning it around, this fact can serve as further evidence that this sporadic group indeed acts on the algebra as a symmetry group.


If the idea described in the last paragraph is true, a mathematical consequence will be the following: the denominator twisted by an element $g\in M_{24}$ should be given by an infinite product involving the $K3$ elliptic genus twisted by $g^k$, or equivalently its Fourier coefficients $\Tr_{K_n^\natural} (g^k)$, with $k$ again running from one to the order of $g$. Furthermore its poles should be given by a product of two copies of the twisted $1/2$-BPS partition function $1/\eta_g$. One can show that this is indeed true for all elements $g\in M_{24}$.
In fact, for the elements $g$ generating the $\Z_{2,3,5,7}$ symplectic automorphisms of $K3$, we will show that the twisted denominator is the same object  considered recently \cite{JatkarJHEP0604:0182006,DavidJHEP0606:0642006} by Sen and his collaborators. 
In these references, they are computed as the generating function of twisted $1/4$-BPS indices \cite{Sen2009,Sen2010}, and are studied as a microscopic test for the quantum entropy function proposal for black holes with $AdS_2$ as part of the near horizon geometry \cite{Sen2008a}.

These observations provide important new insights into the symmetries and structures of the $K3$-compactified string theories, or equivalently toroidally compactified heterotic strings. Furthermore, the assumption of an underlying $M_{24}$ symmetry leads to new conjectural formulas for the twisted $K3$ elliptic genera, the twisted perturbative spectrum of heterotic strings, and the twisted indices for $1/4$-BPS dyons in the ${\cal N}=4$, $d=4$ theories. 

The rest of the paper is organised as follows. 
In section \ref{Twisted Elliptic Genera}, we study the twisted $K3$ elliptic genera and clarify their relations to the McKay-Thompson series of $M_{24}$. From this consideration we give predictions for the twisted $K3$ elliptic genera which have not been computed before. 
In section \ref {The Two Versions of $M_{24}$ Moonshine and the Algebra for BPS Dyons} we start with a review on some previously known mathematical and physical results. 
These include on the one hand the relation between $\eta$-products, $M_{24}$, and heterotic string spectrum, and the relationship between the $K3\times T^2$-compactified string theory and a certain generalised Kac-Moody superalgebra on the other hand. 
We then study the (twisted) denominators of this algebra, and show they relate the two sets of modular objects (elliptic genera and $\eta$-products) which are both attached to $M_{24}$. Finally, in the last section we summarise the lessons learned from this study. In particular we summarise our conjectural formulas for the various (twisted) spectra of the theories arising from $K3$-compactified string theory, while some of the symmetries are yet to be understood explicitly.

\section{Twisted $K3$ Elliptic Genera and the McKay-Thompson Series of $M_{24}$}
\label{Twisted Elliptic Genera}
\setcounter{equation}{0}

In this section we will study the twisted elliptic genera of $K3$ surfaces and discuss their relation to the characters of a certain infinite-dimensional graded representation of the largest Mathieu group $M_{24}$.

In Ref. \cite{Eguchi2010}, Eguchi, Ooguri and Tachikawa studied the following decomposition of the elliptic genus of $K3$ in terms of characters of representations of an ${\cal N}=4$ superconformal algebra
\bea\notag
{\cal Z}_{K3}(\t,z) &=& \Tr_{\text{RR}}\left((-1)^{J_0+\bar J_0} q^{L_0} \bar q^{\bar L_0} e^{2\p i z J_0} \right)\\ \label{K3_elliptic}
&=& \frac{\th_1^2(\t,z)}{\eta^3(\t)}\,\left(\chi \,\m(\t,z) +q^{-1/8}\left(- 2  + T(\t)\right) \right)\;,
\eea
where the first term corresponds to the the short multiplet of spin zero in the Ramond sector, $\chi=24$ is the Euler number of $K3$, and we have written
$$
q=e^{2\p i\t}\quad,\quad y=e^{2\p i z}\;
. 
$$

Notice that the elliptic genus is a weak Jacobi form of the modular group $SL(2,\Z)$, as one can show from the standard CFT spectral flow argument. 
 On the other hand, the so-called Appel-Lerch sum
 $$
 \m(\t,z) = \frac{-i y^{1/2}}{\th_{1}(\t,z)}\,\sum_{n=-\inf}^\inf \frac{(-1)^{n} y^n q^{n(n+1)/2}}{1-y q^n}
 $$
 is not such a simple modular object and is in fact a mock theta function \cite{zwegers,zagier_mock}. This implies that the $q$-series $T(\t)$ does not have simple modular transformation either.  
 The appearance of such a mock modular object in the moonshine for the sporadic group $M_{24}$ is very interesting, though a detailed discussion is out of the scope of the present paper. 

The fascinating observation made by the authors of \cite{Eguchi2010} is about the object $T(\t)$ which shows up in the above superconformal decomposition of the $K3$ elliptic genus (\ref{K3_elliptic}). They observe that the first few Fourier coefficients, which read
\be\label{eot_eqn}
T(\t)= 2(45 q + 231 q^2 + 770 q^3 + 2277 q^4 +5796 q^5 +(3520+10395) q^6 +\dotsi) \;,
\ee
are given by the (sums of) dimensions of irreducible representations of the sporadic group $M_{24}$.

In other words, it is as if there is an infinite-dimensional representation, which we will call $K^\natural$, of the largest Mathieu group, that has a $\Z$-grading related to the $L_0$ of the CFT, such that 
\be\label{eot_eqn2}
K^\natural  = \bigoplus\limits_{n=1}^\inf K^\natural_n\quad,\quad T(\t) = \sum_{n=1}^\inf q^n \, \text{dim}(K^\natural_n)\;.
\ee
Furthermore, the Euler number $\chi =24$ can similarly be written as
$$\chi =\text{dim}(\rho_1\oplus\rho_{23})=1+23\;,$$
where $\r_1$ denotes the trivial representation and $\rho_{23}$ the 23-dimensional representation of $M_{24}$ \cite{Eguchi2010}. 

To further test this proposal and to gain more information about the purported $M_{24}$ module $K^\natural$, a natural set of objects to consider is the twisted version of $K3$ elliptic genus. 
Quite a few things about these twisted elliptic genera have been studied in recent years in the program to understand ${\cal N}=4$, $d=4$ string theory compactifications~\cite{Sen2007a}\nocite{Sen2007b,Cheng2008,Cheng2009,Cheng2007a,Cheng2008a,Dabholkar2008,DavidJHEP0701:0162007,DavidJHEP0611:0732006,DavidJHEP0606:0642006,DavidJHEP0611:0722006,Govindarajan2009,JatkarJHEP0604:0182006,SenAdv.Theor.Math.Phys.9:527-5582005,Sen2010,DabholkarJHEP0711:0772007,Sen2009,Sen2008,Sen2007,SenJHEP0705:0392007,Govindarajan2008,CardosoJHEP0603:0742006,Dabholkar2008a,Dabholkar2008c,Dabholkar2009b,Dabholkar2008b,Banerjee2008,Banerjee2007,Banerjee2008a,Banerjee2008b,Mukherjee2007,DabholkarJHEP0712:0872007,
Mukhi2008a,ShihJHEP0610:0872006}-\cite{Gaiotto2007}.
 These include 
the so-called CHL models in heterotic string theory \cite{Chaudhuri1995}, or
type II string theory compactified on $K3\times T^2$ orbifolded by a $\Z_N$ symmetry which acts on $K3$ as an order $N$ symplectic automorphism and as a $\Z_N$ shift along one of the circles in $T^2$. 
 In particular,  the twisted elliptic genera has been computed for all $\Z_p$ symmetries with prime $p$ in the Nikulin's list of $K3$ quotients  \cite{DavidJHEP0606:0642006}: letting $g_p$ be the order $p$ generator  of the discrete $\Z_p$ transformation on the geometry of $K3$, the  following object in the $K3$ CFT was considered by the authors of \cite{DavidJHEP0606:0642006} 
\be\label{twisted_elliptic_genus}
  \Tr_{\text{RR},(g_p)^m}\left((-1)^{J_0+\bar J_0} (g_p)^n q^{L_0} \bar q^{\bar L_0} e^{2\p i z J_0} \right)\;,\; m,n = 0,\dotsi,p-1\;,
\ee
where the trace is taken over the $(g_p)^m$-twisted sector. 
We are interested in the case $m=0,n=1$, the answer for which is given by
 \bea\notag
 {\cal Z}_{g_p}(\t,z)&=&   \Tr_{\text{RR}}\left((-1)^{J_0+\bar J_0} g_p \,q^{L_0} \bar q^{\bar L_0} e^{2\p i z J_0} \right)\\ \notag
 &=&  \frac{2}{p+1}\f_{0,1}(\t)+ \frac{2p}{p+1}\f_{-2,1}(\t,z)\f_2^{(p)}(\t)\\\label{prime_N_answer} 
 &&\text{for }(g_p)^p={\mathds 1}\;,\;p=2,3,5,7\;.
 \eea
where $\f_{-2,1}(\t)$ is the weight $-2$, index one weak Jacobi form and $\f_2^{(p)}(\t)$
is a weight two modular form under the congruence subgroup $\G_0(p)\subset PSL(2,\Z)$. The total expression ${\cal Z}_{g_p}(\t,z)$ hence transforms as a weak Jacobi form of weight zero, index one of $\G_0(p)$. More details on these modular objects can be found in Appendix \ref{Modular Stuff}. 
 
For $\Z_N$ orbifolds with $N$ not prime, ideas for how to compute the twisted elliptic genus using geometric data have been put forth in 
\cite{Govindarajan2009,Sen2010}, as well as an explicit formula for the case $N=4$ in \cite{Govindarajan2009}. 

As was mentioned earlier, it is known that all the groups in the Nikulin's list of $K3$ quotients can be embedded in $M_{24}$ in a natural way. 
Provided that this ``classical geometric" part of $M_{24}$ is indeed realised  
 on the purported representation $K^\natural$ according to the embedding of Mukai and Kondo, 
it is a mere consequence of the conjecture (\ref{eot_eqn2}) that, for an element $g$ generating a certain symplectic automorphism, the twisted elliptic genera
\be
{\cal Z}_{g}(\t,z) = \Tr_{\text{RR}}\left((-1)^{J_0+\bar J_0} \,g\, q^{L_0} \bar q^{\bar L_0} e^{2\p i z J_0} \right)
\ee
are related in the following way to the so-called McKay-Thompson series $T_g(\t)$, or twisted partition functions, of $M_{24}$
\bea\label{guess1}
{\cal Z}_{g}(\t,z) &=&  \frac{\th_1^2(\t,z)}{\eta^3(\t)}\,\left(\chi(g)\, \m(\t,z) +q^{-1/8}\left(- 2  + T_g(\t)\right) \right)\\
T_g(\t)&=& \sum_{n=1}^\inf (\Tr_{K^\natural_n}g)\; q^n\quad,\quad g\in M_{24}\;.
\eea
In the above formula, $\chi(g)={\cal Z}_{g}(\t,z=0)$ is the twisted Euler number (trace of $g$ on $K3$ cohomologies) which is expected to be
\be\label{twisted_euler}
\chi(g)= \Tr_{\r_1\oplus \r_{23}}g\;.
\ee

A few comments are in order here. First, taking the identity element $g={\mathds{1}}$, we recover the series $\sum_{n=1}^\inf \text{dim}(K^\natural_n) q^n = T(\t)$ encoding the dimensions of the representation.  Second, it is clear from their definition that the McKay-Thompson series depend only on the conjugacy class of the group element. Namely, we have 
$$
T_g(\t) = T_{hgh^{-1}}(\t)\;.
$$
Despite of the large number  $$|M_{24}|=2^{10}\cdot 3^3\cdot 5\cdot 7\cdot 11\cdot 23\sim 10^8$$ of elements, $M_{24}$ has only $26$ conjugacy classes, as listed in Table \ref{M24}. Hence we expect at most $26$ distinct McKay-Thompson series $T_g(\t)$. Third, the modularity properties of the above McKay-Thompson series $T_g(\tau)$ are somewhat obscure due to the subtraction of the Mock theta function $\m(\t,z)$ and 
 are most easily seen in the combined CFT object ${\cal Z}_g(\t,z)$. This is to be contrasted with the more 
familiar case of the Monstrous Moonshine, where the corresponding McKay-Thompson series are all modular functions of discrete subgroups of $PSL(2,\R)$. 

Finally, following the same logic we also expect the more general twisted elliptic genus which can be calculated when the CFT has a symmetry group $G\subset M_{24}$
\be{\cal Z}_{(h,g)}(\t,z) =\Tr_{\text{RR},h}\left((-1)^{J_0+\bar J_0} \,g\, q^{L_0} \bar q^{\bar L_0} e^{2\p i z J_0} \right)\quad,\quad h, g\in G\;,\ee 
to be related to the so-called generalised McKay-Thompson series for commuting pairs of elements $(g,h)$\cite{generalized_moonshine}
\be\label{generalisedMcKayT}
T_{h,g} (\t)=   \sum_{n=1}^\inf (\Tr_{K^\natural_n,h}g)\; q^n\quad,\quad g,h\in M_{24}
\ee
of the group $M_{24}$, where the element $h$ denotes the twisted sectors in the Hilbert space. 

Now we are ready to test the proposal (\ref{guess1}) for the known cases when $g$ is taken to be the element generating the $\Z_{p=2,3,5,7}$ symplectic automorphisms of $K3$. Rewriting the known answer (\ref{prime_N_answer}) we get the results that the twisted elliptic genus ${\cal Z}_{g_p}(\tau)$ can indeed be written in the form (\ref{guess1}). The corresponding elements $g_p\in M_{24}$ in their ATLAS names and the first seven Fourier coefficients of the $q$-series $T_{g_p}(\tau)$ are given in Table \ref{prime_p_coefficients_table}.

 \begin{table} \centering \resizebox{1\textwidth}{!}{
 \begin{tabular}{c|c|ccccccc}
 \toprule
 $g$&$\chi(g)$&$q^1$&$q^2$&$q^3$&$q^4$&$q^5$&$q^6$&$q^7$\\\cline{3-9}
$1A$&24& 2$\times$45&2$\times$231&2$\times$770&2$\times$2277&2$\times$5796&2$\times$13915&2$\times$30843\\
 $2A$&8&2$\times$-3&2$\times$7&2$\times$-14&2$\times$21&2$\times$-28&2$\times$ 43&2$\times$-69\\
 $3A$&6&0&2$\times$-3&2$\times$5&0&2$\times$-9&2$\times$10&0\\
 $5A$&4&0&2$\times$1&0&2$\times$-3&2$\times$1&0&2$\times$7\\
 $7A$&3&-1&0&0&4&0&-2&2\\
 \bottomrule
  \end{tabular}}
  \caption{ \label{prime_p_coefficients_table}\footnotesize{Here we list the first seven coefficients of the $q$-series $T_{g}(\tau)$ that show up in the elliptic genera of $K3$ twisted by $\Z_p$ symplectic automorphisms (\ref{prime_N_answer},\ref{guess1}), together with the twisted Euler number. In the first column we have adopted the same ATLAS naming for the conjugacy classes as in Table \ref{M24}. In particular, in the first row are the coefficients of the non-twisted $K3$ elliptic genus as given in \cite{Eguchi2010}. }}
 \end{table}
 
Comparing this coefficient table with the character table of $M_{24}$ (Table \ref{M24}), there are quite a few things to be learned. 
First of all, 
the first six columns are consistent with the hypothesis that $K_{i}^\natural, i=1,\dotsi,5$ are given by the irreducible representations of $M_{24}$ with the corresponding dimensions. Also the twisted Euler number 
$\chi(g_p) = \frac{24}{p+1}$ is consistent with the formula (\ref{twisted_euler}).
This gives very strong new evidence for the possibility that $M_{24}$ does act on the elliptic genus of $K3$. Furthermore, this also shows that the part of $M_{24}$ that does have interpretation as automorphisms of classical $K3$ geometry indeed acts on $K3$ in a way that we expect it to. 

Second, we see that the McKay-Thompson series give useful extra information about the $M_{24}$ module $K^{\natural}$. Notice that the dimensions of irrep's of $M_{24}$ are highly degenerate. For $n>5$, $K_n^{\natural}$ are not simply given by irreducible representations anymore but rather by positive integral combinations of them. The decompositions are in fact never unique and one needs more information than the dimension to determine the representation  $K_n^{\natural}$. For example, for $n=7$ the authors of \cite{Eguchi2010} propose 
$$
30843 = 10395+2\times 5796+5544+3312\;,
$$
which is unfortunately inconsistent with the Fourier coefficients of the McKay-Thompson series listed in Table \ref {prime_p_coefficients_table}. 

Instead, the following decomposition for example, is consistent with the limited character table we have so far and also the examples we are going to consider later\footnote{It is in fact the only possibility if one assumes that the largest irrep appears once, irrep's with dimension smaller than $500$ do not appear, and each irrep appears at most twice. Similar computer scans can be extended to more general assumptions. }
$$
30843 = 10395+ 5796+5544+5313+2024+1771\;.
$$
This is a concrete example of how the twisted $q$-series contain much more information about the appropriate representation. We expect them to give important hints about the representation before we eventually obtain other ways to ``naturally" define  $K^\natural$ from physical considerations. 

Finally, it is a familiar fact that all Fourier coefficients of $K3$ elliptic genus are even integers. As can be seen in the table and can be shown in general \cite{Cheng2008}, this remains true for the $K3$ elliptic genus twisted by element generating the $\Z_{2,3,5}$ automorphisms.  A directly related statement about the generalised Kac-Moody alegebra that generates the $1/4$-BPS spectrum  in the corresponding ${\cal N}=4$, $d=4$ string theory will be discussed in section \ref{The BPS Dyon Algebra and the Two Moonshines}. Notice though that in the last row in Table \ref{prime_p_coefficients_table}, the Fourier coefficients are no longer even for the $\Z_7$ twisted character. Furthermore, by comparing with the character table we conclude that all representations in $K^{\natural}$ come  in  conjugate pairs\footnote{We say that an irreducible representation is its own conjugate if it does not come in conjugate pairs.}. In other words, a better way to write the numbers in Table \ref{prime_p_coefficients_table} is ${\bf c}+\overline{{\bf c}}$ instead of $2\times {\bf c}$. Such a distinction is invisible unless one studies the twisted version of the partition functions of the conformal field theory. Later we will see how this sheds new light on the structure of $1/4$-BPS spectrum of the $K3\times T^2$-compactified string theory. 

After this simple exercise we are now not only more confident about the existence of the $M_{24}$ module $K^\natural=\oplus_n K^\natural_{n}$, but also how the few lowest-lying representations $K^\natural_{\;n}$ look like. 
Combined with modularity properties, the proposal (\ref{guess1}) can be used to derive explicit expressions for the $K3$  elliptic genera twisted by various elements of $M_{24}$. This derivation is purely arithmetic and does not depend on whether the element $g\in M_{24}$ has a classical geometrical meaning or not.  To illustrate this, let us first derive the $K3$ elliptic genera  twisted by elements that generate the $\Z_{4,6,8}$ symplectic automorphisms of $K3$. Together with the $\Z_{p=2,3,5,7}$ we just discussed, they exhaust the $\Z_N$ groups in the list of Nikulin's $K3$ involutions. An alternative proposal to compute these $\Z_{4,6,8}$-twisted elliptic genera from geometric consideration can be found in \cite{Sen2010}.


From the standard CFT arguments about the change of boundary conditions under modular transformations, we expect ${\cal Z}_{g}(\t,z)$ to be invariant up to a phase under transformations
$$
\t \to \frac{a \t+b}{c\t +d}\quad,\quad z\to \frac{z}{c\t+d} \quad,\quad \bigg[\!\begin{array}{cc} a& b\\c&d\end{array}\!\bigg] \in SL(2,\Z)\;,\; c = 0 \;\text{mod  ord}(g) \;.
$$
This condition leaves just a few unknown coefficients to determine in the expression for ${\cal Z}_{g}(\t,z)$. See Appendix \ref{Modular Stuff} for more information on the relevant modular objects. 

From this consideration, we obtain 
\bea\notag
{\cal Z}_{4B}(\t,z)&=& \frac{1}{3}\f_{0,1}(\t)+ \f_{-2,1}(\t,z)\left(-\frac{1}{3}\f_2^{(2)}(\t)+2\f_2^{(4)}(\t)\right)\\ \notag
{\cal Z}_{6A}(\t,z)&=& \frac{1}{6}\f_{0,1}(\t)+ \f_{-2,1}(\t,z)\left(-\frac{1}{6}\f_2^{(2)}(\t)-\frac{1}{2}\f_2^{(3)}(\t)+\frac{5}{2}\f_2^{(6)}(\t)\right)\\\label{N468_elliptic_genera}
{\cal Z}_{8A}(\t,z)&=& \frac{1}{6}\f_{0,1}(\t)+ \f_{-2,1}(\t,z)\left(-\frac{1}{2}\f_2^{(4)}(\t)+\frac{7}{3}\f_2^{(8)}(\t)\right)\;.
\eea

To make our discussion unambiguous, in the above equation we have adopted the ATLAS naming of the conjugacy classes, as listed in Table \ref{M24}. 
Comparing the answers to the known result, ${\cal Z}_{4B}(\t,z)$ is identical to the expression given in Ref.~\cite{Govindarajan2009} for $\Z_4$-twisting. 
For twisting by $\Z_6$ and $\Z_8$ geometric isometries, there are no explicit formulas available in the literature, and our answer is consistent with the topological constraints in \cite{Govindarajan2009}. Furthermore, for $N={2,3,4,6}$, we observe that our results can be rewritten as 
\be
\chi(g_N)\; \frac{\th_1(\t,z+\frac{1}{N})\th_1(\t,-z+\frac{1}{N})}{\th_1^2(\t,\frac{1}{N})}\;,
\ee
which can be interpreted geometrically as suggested in \cite{Sen2010}.
In the above formula, $\chi(g_N)=\Tr_{\r_1\oplus \r_{23}}g_N$ denotes the Euler number twisted by the order $N$ elements $2A,3A,4B$ and $6A$ ($N=2,3,4,6$) respectively. 

This procedure can be extended to various other elements of $M_{24}$, even when they do not have an interpretation as holonomy-preserving automorphisms of $K3$. We expect these twisted elliptic genera to provide valuable information about the action of the corresponding elements on stringy $K3$ geometry.  For the brevity of the text, the formulas of these other twisted elliptic genera are placed in Appendix \ref{Formulas for Twisted Elliptic Genera}.

\section{The Two Versions of $M_{24}$ Moonshine and the Algebra for BPS Dyons}
\label{The Two Versions of $M_{24}$ Moonshine and the Algebra for BPS Dyons}
\setcounter{equation}{0}

In this section we will first review the known relationship between products of $\eta$-functions, the perturbative spectra of heterotic string compactifications, and the sporadic group $M_{24}$. Putting different pieces of known results together, we observe how an old and the new versions of ``$M_{24}$ moonshine" are related in a generalised Kac-Moody algebra, which is known to have the physical interpretation as the spectrum-generating algebra for $1/4$-BPS dyons in the $K3\times T^2$-compactified type II string theory.

\subsection{Review: $\eta$-products, $M_{24}$, and Heterotic Strings}
\label{Review}
It is a familiar fact that the $1/2$-BPS partition function of heterotic string compactified on $T^6$ is nothing but the bosonic string partition function
$$
\frac{1}{\eta^{24}(\t)} = q^{-1}+\sum_{n=0}^\inf d_n q^n\quad,\quad q= e^{2\p i\t}\;.
$$

In \cite{SenAdv.Theor.Math.Phys.9:527-5582005,JatkarJHEP0604:0182006}, the $1/2$-BPS partition functions of the above theory twisted by $\Z_{p=2,3,5,7}$ symmetries have been computed to be
\be\label{perturbative_string_prime}
\frac{1}{\eta^{\frac{24}{p+1}}(\t)\eta^{\frac{24}{p+1}}(p\t)}\;.
\ee

Very recently, Govindarajan and Krishna \cite{Govindarajan2009}
extended the above results  to other $\Z_N$ orbifolds, with $N=4,6,8$ being the non-prime numbers in the list of $\Z_N$ quotients in Nikulin's list of $K3$ quotients. 
The fact that all $K3$ symplectic automorphisms can be embedded in $M_{23}$ and hence in $M_{24}$ was crucial for their derivation. To understand the answer we have to know a few more things about  $M_{24}$. The largest Mathieu group can be thought of as a subgroup of the group $S_{24}$ of the permutation of $24$ objects. In this way, one can assign to each conjugacy class of $M_{24}$ a corresponding ``cycle shape" of the form $$
1^{i_1}2^{i_2}\dotsi r^{i_r}\quad,\quad \sum_{\ell=1}^r \ell \,i_\ell =24\;.$$
See Table \ref{M24} for the cycle shapes of all conjugacy classes.  
 Using Mukai's embedding of $K3$ symplectic automorphisms into $M_{23}\subset M_{24}\subset S_{24}$, one can also associate to a generator of an $K3$ symplectic automorphism a cycle shape. For example, the cycle shape of the identity element is $1^{24}$, which denotes the collection of $24$ cycles of length one. Similarly, the generator of the geometric $\Z_2$ symplectic automorphism is associated with $1^82^8$, the collection of $8$ cycles of length $1$ and $8$ cycles of length $2$, and $\Z_3$ is given by $1^6 3^6$ and so on. In the simple cases, the explicit meaning of the cycle shapes in terms of Narain lattices with enhanced symmetries on the heterotic side can be found in \cite{Chaudhuri1996}. 
 
Using this mathematical relation to the Mathieu group and the asymptotic growth of the degeneracies in the physical system, it was argued in \cite{Govindarajan2009} that the result (\ref{perturbative_string_prime}) can be extended to $\Z_N$ orbifolds with composite $N$, and the answer is given by the corresponding $\eta$-products. To be more specific, through the embedding of $K3$ automorphisms into $M_{24}$, the $\Z_4$ automorphism corresponds to the cycle shape $1^4 2^2 4^4$ and the corresponding twisted generating function is given by $(\eta^{4}(\t)\eta^2(2\t)\eta^4(4\t))^{-1}$, and the $\Z_6$ theory has $(\eta^{2}(\t)\eta^2(2\t)\eta^2(3\t)\eta^2(6\t))^{-1}$, the $\Z_8$ theory $(\eta^{2}(\t)\eta(2\t)\eta(4\t)\eta^2(8\t))^{-1}$. For $\Z_p$ with $p$ being prime, this procedure reproduces the known result (\ref{perturbative_string_prime}).

In the same way, one can associate a so-called $\eta$-product to every conjugacy class of $M_{24}$, whether it has a known interpretation as a classical $K3$ automorphism or not.
In other words, we have a map between all elements $g\in M_{24}$ and cusp forms $\eta_g(\t)$
\be\label{eta_prod}
g \text{ with cycle shape }1^{i_1}2^{i_2}\dotsi r^{i_r} \mapsto \eta_g(\t) = \prod_{\ell =1}^r \eta(\ell \t)^{i_\ell}=\sum_{m=1}^\inf a_{g,m} q^m
 \ee

These $\eta$-products are long known to be the central objects of a version of ``moonshine" of the group $M_{24}$ \cite{Mason,mason_2}. In other words, the Fourier coefficients of the of these  $\eta$-products given by the cycle shapes of $M_{24}$ are known to be the virtual, or generalised, characters of $M_{24}$. That is, the Fourier coefficients of $\eta_g(\t)$ are (not necessarily positive) integral linear combinations of irreducible characters of $M_{24}$. 
For example, we have 
\bea\notag
\eta^{24}(\t)  &=& q + (-1-23)q^2 +(-1+253)q^3 + \,\dotsi \\ \notag
\eta^8(\t) \eta^8(2\tau) &=& q + (-1-7) q^2 + (-1+13) q^3+\,\dotsi\\ \notag
\eta^6(\t)\eta^6(3\t) &=& q + (-1-5)q^2 + (-1+10) q^3 +\,\dotsi \\
\dotsi 
\eea
Interested readers can see \cite{RamanunjanEta} for a list of decompositions in terms of irrep's of $M_{24}$ from $q^1$ to $q^{30}$.

For completeness we also mention here the second formulation of the relationship between $\eta$-products and $M_{24}$. 
Using the remarkable fact that all the $\eta$-products obtained from (\ref{eta_prod}) are multiplicative (a Fourier series $F(\tau)=\sum_{m=1}^\inf a_m q^m$ is called multiplicative if $a_1 a_{mn}=a_m a_n$ for $(m,n)=1$), one can show that its Mellin transform 
$$
F_g(s) = \sum_{m=1}^\inf a_{g,m} m^{-s}
$$
is the Euler product $\prod_p (1- a_{g,p} p^{-s} + b_{g,p} p^{-2s})$, where $b_{g,p} =\e_{g,p} p^{k-1}$ with $\e_{g,p}=0,\pm1$ certain Dirichlet character and $k$ is the so-called ``level" of the $M_{24}$ element $g$, defined as half of the total number of cycles or equivalently the weight of the $\eta$-product $\eta_g(\t)$. Then the statement is, for all prime $p$ with $p\neq 3$, $b_{g,p}$ is a character of $M_{24}$. For example, for $p=2$ we have 
\bea\notag
1\times2^{12-1} &=& 1+23+253+1771\\ \notag
0\times 2^{8-1}&=& 1 + 7 + 13 - 21 \\ \notag
1\times2^{6-1} &=& 1+5+10+16\\ \notag
&&\\ \notag
\dotsi
\eea
for $g=1A,2A,3A,\dotsi$.

\subsection{The BPS Dyon Algebra and the Two Moonshines}
\label{The BPS Dyon Algebra and the Two Moonshines}
It is a rather curious fact that we seem to have now two sets of modular objects, ${\cal Z}_g(\t,z)$ and $\eta_g(\t)$, each giving a version of moonshine for the group $M_{24}$. Notice also that the underlying $M_{24}$ module for these two sets of objects seem to be completely unrelated to each other.  But in this subsection we would like to point out that they are in fact united in the denominator of the characters of a generalised Kac-Moody superalgebra, which has the physical interpretation as the spectrum-generating algebra for the $1/4$-BPS states in the $K3\times T^2$-compactified type II string theory, or equivalently the $T^6$-compactified heterotic string theory.

To explain this, let us remind ourselves a few things about generalised Kac-Moody superalgebras. See, for example, \cite{Grit_Nik,ray_book,Cheng2008a} for a bit more or much more information.  A generalised Kac-Moody superalgebra, or sometimes called a Borcherds-Kac-Moody algebra, is an infinite dimensional algebra with non-positive definite Cartan matrix (Kac-Moody).  Furthermore it contains the so-called imaginary simple roots (generalised), which are simple roots which are time- or light-like (having positive or null norm squared in our convention). Finally the roots come in bosonic or fermionic  varieties (super). A very important object to consider in these algebras is the denominator of their characters, which satisfies the so-called denominator identity that relates an infinite sum to an infinite product: 
\be\label{denominator}
e^\varrho \prod_{\a\in \D_+} (1-e^\a)^{\text{mult}(\a)} = \sum_{w\in W} \e(w)\,w(e^\varrho \Sigma) \;.
\ee
Here $\D_+$ denotes the set of positive roots and $\text{mult}(\a)$ the root multiplicity, $\varrho$ the Weyl vector of the algebra satisfying the condition $(\varrho,\a_i)=(\a_i,\a_i)/2$ for all real simple roots (simple roots that are space-like) $\a_i$, $W$ the Weyl group which is typically a hyperbolic reflection group and $\e(w)=\pm1$. Finally, $\Sigma$ is some combination of simple imaginary roots which encodes the information of the degeneracies of these time-or light-like simple roots. 

More than a decade ago a generating function has been proposed for the $1/4$-BPS spectrum of the ${\cal N}=4$, $d=4$ string theory obtained by compactifying type II strings on $K3\times T^2$ \cite{DVV}. There it was also observed that the generating function has the form of the square of the denominator of a generalised Kac-Moody superalgebra. This algebra was first constructed by Gritsenko and Nikulin in \cite{Grit_Nik} and has a $(2,1)$-dimensional Cartan matrix 
$$
\left(\!\begin{array}{rrr} 2&-2&-2\\ -2&2&-2\\ -2&-2&2\end{array}\!\right)
$$
for the real roots. It is also the Gram matrix with entries $-2(\a_i,\a_j)$ in our metric convention.
Later in \cite{Cheng2008a}, a more physical interpretation is given to the presence of this algebra. In particular, its Weyl group is identified with the group controlling the wall-crossing phenomenon of the physical spectrum. 

As is often the case for automorphic forms arising as the denominator of generalised Kac-Moody algebras, both the Fourier coefficients in the sum side and the exponents on the product side of the formula (\ref{denominator}) are given by the Fourier coefficients of some (typically different) modular objects. These phenomena are  often referred to as the ``arithmetic lift" and the ``Borcherds lift" respectively. 
For the algebra of interest here, the two sides of the (square of the) denominator formula (\ref{denominator}) is given by  
\bea\notag
\F(\O) &=& e^{4 \p i (\varrho,\O)} \prod_{\substack {\a}} \left( 1- e^{2\p i (\a,\O)}\right)^{c(|\a|^2)} \\  \label{phi10}
&=& \sum_{m>0} \;e^{2\p i m \s} \,T_m \f_{10,1}(\r,\n)\;,\\\notag
 \eea
where $T_m$ denotes the $m$-th Hecke operator and we write the complexified vector in the $(2,1)$-dimensional weight space in a matrix form as
$$
\O= \bem\r & \n\\\n&\s\eem \;,
$$
with the norm $|X|^2= (X,X)=\text{det}X$. The Weyl vector is given in terms of the three real simple roots by $\varrho=\frac{1}{2}\sum_{i=1}^3 \a_i$, and a choice for the set of real simple roots is 
$$
\a_1 =\bem0&-1\\-1&0\eem\;,\;\a_2 =\bem2&1\\1&0\eem\;,\;\a_3 =\bem0&1\\1&2\eem\;.
$$
The product is then taken over $\a$'s that are positive semi-definite integral linear combinations of the $\a_{1,2,3}$:
$$
\a\in\{\Z_+ \a_1+\Z_+ \a_2+\Z_+ \a_3\}\;.
$$
Notice that the denominator $\F(\O)$ is independent of the choice of real simple roots, or equivalently a choice of the fundamental Weyl chamber, as can be seen explicitly from the sum side.

The exponents (root multiplicities) $c(k)$ on the product side and the Fourier coefficients on the sum side are then given by the weak Jacobi forms
\bea\notag
Z_{K3}(\t,z) = 2 \f_{0,1}(\t,z) &=& \sum_{n,\ell\in\Z,n>0}  \,c(4n-\ell^2) q^n y^\ell\\
\f_{10,1}(\t,z)&=& 
 -\frac{\th_{1}^2(\t,z)}{\eta^6(\t)}\,\eta^{24}(\t) 
 \;,
\eea
and $\F(\O)$ is an automorphic form of the genus two modular group $Sp(2,\Z)$ of weight $10$. In the last interpretation, $\O$ should really be thought of as the period matrix of the genus two surface.

The $1/4$-BPS index with a given charge $(P,Q)\in \G^{22,6}\oplus\G^{22,6}$ and at a given point in the moduli space is then determined in the following way. Consider the following two vectors in the future light-cone of $\R^{2,1}$ 
\bea\notag
\L_{(P,Q)} &=& \bem Q\cdot Q & Q\cdot P \\ Q\cdot P& P\cdot P \eem\quad\text{and}\\\notag\, \til Z &=& \frac{1}{\l_2}\bem |\l|^2 & \l_1 \\ \l_1 &1\eem + \frac{1}{\sqrt{Q_R^2 P_R^2 - (Q_R\cdot P_R)^2}}\bem Q_R\cdot Q_R & Q_R\cdot P_R \\ Q_R\cdot P_R& P_R\cdot P_R \eem\;,
\eea
where $\l=\l_1+i\l_2$ is the heterotic axion-dilaton and $Q_R$, $P_R$ are the right-moving part of the charges. See \cite{Cheng2008a} for the details and motivations for these formulas.  
There are unique elements $w_1$, $w_2$ of the Weyl group such that both $w_1(\L_{(P,Q)})$ and $w_1w_2(\til Z)$ lie inside the fundamental Weyl chamber, which is defined as the region bounded by the planes of orthogonality to the real simple roots $\a_{1,2,3}$. 
Choosing the set of real simple roots $\a_{1,2,3}$ such that $w_1={\mathds{1}}$, the corresponding $1/4$-BPS index is then given by $D(w_2(\L_{P,Q}))$, defined by
\bea
\frac{1}{\F(\O)}  = \sum_{\L\in {\cal W}, w\in W} D(w(\L)) e^{2\p i (w(\L),\O)}\;.
\eea
In the above formula, an expansion in $e^{2\p i (\a_i,\O)}, i=1,2,3$ on the left-hand side should be understood. See also \cite{Cheng2007a} for an equivalent formula in terms of a contour integral with moduli-dependent contours. 

As mentioned earlier, the Weyl group of the algebra plays an important role in the counting of dyons due to the so-called wall-crossing phenomenon in this theory. When the moduli-vector $\til Z$ hits the wall of a Weyl chamber, some $1/4$-BPS bound states corresponding to bound states of $1/2$-BPS particles appear or disappear from the spectrum and as a result the $1/4$-BPS index jumps. This physical phenomenon is reflected in the following property of the partition function \cite{SenJHEP0705:0392007}. The denominator $1/\F(\O)$ has double poles at the location where the vector $\O$ lies on wall of the Weyl chamber and factorises into two parts, each corresponds to the $1/2$-BPS partition function. For instance, 
\be
\lim_{\n\to 0} \frac{1}{\F(\O)} = -\frac{1}{4\p^2 \n^2} \frac{1}{\eta^{24}(\r)} \frac{1}{\eta^{24}(\s)}\;,
\ee
and similarly for all other poles.

Apart from the $1/4$-BPS index, or the sixth helicity supertrace to be more precise, finer information about the spectrum can be obtained when extra isometries are present. Recently, such ``twisted $1/4$-BPS indices"
$$
B_g(Q,P) = \frac{1}{6!} \text{Tr}_{{\cal H}_{Q,P}} \left( g (-1)^{2J_3} (2J_3)^6 \right)  
$$
of the above theory have been studied \cite{Sen2009,Sen2010} when there is an enhanced geometric $\Z_{p=2,3,5,7}$ symmetry in the internal manifold $K3\times T^2$. Here $J_3$ denotes the third component of the spacetime angular momentum quantum number. Denoting the generator of the $\Z_{p=2,3,5,7}$ isometry by $g_p$, the corresponding generating function is given by $1/\F_{g_p}(\O)$, where $\F_{g_p}(\O)$ is an automorphic form of a subgroup of $Sp(2,\Z)$ which also has an infinite-product expression\footnote{In fact they are also known to have the following infinite sum expression \cite{JatkarJHEP0604:0182006} 
$$
\F_{g_p}(\O) = \sum_{m>0} \;e^{2\p i m \s} \,T_m \left( -\frac{\th^2_1(\r,\n)}{\eta^6(\r)} \eta_{g_p}(\r) \right)\quad,\quad p=2,3,5,7
$$
just as in the untwisted case (\ref{phi10}). Unlike in the Monster case, the assumption that the sporadic group $M_{24}$ is a symmetry of the positive root system does not directly imply that it acts on the system of simple roots in a simple way, and hence it is not straightforward to twist the sum side of the denominator identity.  From the form of the above equation it is nevertheless tempting to see whether it is possible to generalise it to other elements of  $M_{24}$. As to be expected probably, it fails for all conjugacy classes of $M_{24}$ for which the object $ -\frac{\th^2_1(\r,\n)}{\eta^6(\r)} \eta_{g}(\r)$ has zero or negative weight, and furthermore does not seem to hold for the conjugacy class $3B$. For all other cases it appears to hold up to the first few dozens of coefficients that we have checked.}
 \cite{Sen2009,Sen2010,DavidJHEP0606:0642006}
\bea\notag
\F_{g_p}(\O) = e^{4\p i(\varrho,\O)} \prod_{r=0}^{p-1} \prod_{\a}\left( 1- e^{2\p i (r/p)} e^{2\p i(\a,\O)}\right)^{\frac{1}{p}\sum_{s=0}^{p-1} e^{-2\p i r s/p}c_{(g_p)^s}(|\a|^2) } \\\label{twisted_den1}
\eea
In this formula, 
the product is again taken over positive roots $\a$ as before, and the exponents $c_{(g_p)^s}(|\a|^2)$ are now the Fourier coefficients of the twisted  $K3$ elliptic genus ${\cal Z}_{(g_p)^s}(\t,z)$ (\ref{twisted_elliptic_genus}).

Now, notice that (\ref{twisted_den1}) can be written as
\be \label{twisted_denominator}
\F_{g}(\O)  = e^{4\p i(\varrho,\O)} \prod_{\a}\exp\left(\sum_{k=1}^\inf -\frac{1}{k} c_{(g)^k}(\lvert\a\rvert^2) \,e^{2\p i(k\a,\O)} \right)\;.
\ee
We recognise this is nothing but the twisted denominator of the original generalised Kac-Moody algebra (\ref{phi10}). 
More precisely, this is the expression one obtains if we regard the subalgebra $E$ spanned by positive roots as an infinite-dimensional representation of $M_{24}$ graded by a (2,1) lattice, and the denominator as calculating the alternating sum of exterior powers $$\L(E)=\L^0(E)\oplus\L^1(E) \oplus\dotsi$$ of $E$ \cite{borcherds_monstrous} . 
Furthermore, once accepting this interpretation, the above formula defines a twisted denominator for all  elements of $g\in M_{24}$, whether or not they are known to generate the geometric $\Z_N$ actions.

To see the appearance of the $\eta$-products in the twisted denominators, let us again look at the wall-crossing poles of the (twisted) partition function which should relate the jump in $1/4$-BPS index to $1/2$-BPS indices. Using 
$$
\sum_{\ell} c_g(4n-\ell^2) = \Tr_{\r_1\oplus \r_{23}}(g) \d_{n,0}\, \;,\; c_g(-1)=-2\;,\; c_g(-n)=0 \text{ for }n>1
$$
one indeed gets
\be
\lim_{\n\to 0 }\frac{1}{\F_g(\O)}  = -\frac{1}{4\p^2 \n^2} \frac{1}{\eta_g(\r)}\frac{1}{\eta_g(\s)}
\ee
for all $g\in M_{24}$.

Now we see that the two versions of moonshine discussed in Section \ref{Twisted Elliptic Genera} and Section \ref {Review}, which involve two seemingly unrelated $M_{24}$ modules, indeed co-exist and get related in the denominator of 
the generalised Kac-Moody algebra which has been studied in the context of dyon counting in the ${\cal N}=4$, $d=4$ theory. 
From the physics point of view, this is consistent with the tentative interpretation we give to $1/\F_g(\O)$ as the generating function of the twisted $1/4$-BPS index, and $1/\eta_g(\t)$ as the twisted $1/2$-BPS partition function, and the wall-crossing formulas that we already know \cite{SenJHEP0705:0392007,Cheng2007a,Cheng2008a}.

This insight about the sporadic symmetry of the dyon algebra does not only shed light on the structure of the non-perturbative structure in the $d=4$ theories, they also help us to predict various (twisted) spectra. We now proceed to discuss these consequences in our last section.

\section{Conclusions and Discussions}
\label{Conclusions and Discussions}
\setcounter{equation}{0}

In this paper, by studying the twisted elliptic genera of $K3$ we first provided further evidence for the idea that associated to the elliptic genus of $K3$ is an infinite-dimensional graded representation of the sporadic group $M_{24}$. After that  we pointed out that this $M_{24}$ module and the previously observed $M_{24}$ moonshine given by the $\eta$-products are in fact united in the denominator of the generalised Kac-Moody algebra that is conjectured to generate the spectrum of $K3\times T^2$-compactified type II string theory. 

From these observations and again using the embedding of the symplectic automorphisms of $K3$ into $M_{24}$, we have the following conservative version of a conjecture: 
for all the symplectic automorphisms of $K3$ generated by $g\in M_{24}$

\begin{enumerate}
\item{At a point in the moduli space where there is an enhanced symmetry generated by $g$, let us consider the $K3$ elliptic genus twisted by $g$. Then this object is given by (\ref{guess1}),
 where $K^\natural$ is an infinite-dimensional graded representation of $M_{24}$ with the first five $K_{n=1,\dotsi,5}^\natural$ given by conjugate pairs of irreducible representations of  $M_{24}$  with the dimensions  indicated in Table \ref{prime_p_coefficients_table}. 
} 
\item{Consider the ${\cal N}=4, d=4$ theory obtained by compactifying type II string theory on $K3\times T^2$ with enhanced symmetry generated by $\til g$ ,which acts as
 a symplectic automorphism generated by $g$ on $K3$ accompanied by a shift action on $T^2$. 
 This shift is uniquely determined up to isomorphisms and we refer to \cite{Chaudhuri1996} for their explicit forms. 
The $1/2$-BPS partition function twisted by such a symmetry 
 is given by $1/\eta_g(\t)$ (\ref{eta_prod}). See also \cite{Govindarajan2009}. }
\item{The $1/4$-BPS spectrum twisted by the same symmetry  is given by the automorphic form $1/\F_g(\O)$, where $\F_g({\O})$ is the twisted denominator defined in the form of an infinite product as in (\ref{twisted_denominator}).}
\end{enumerate}

A bolder, and perhaps more natural, version of the conjecture will state that the full $M_{24}$ acts on moduli-invariant objects in $K3$-compactified string theories, including the $K3$ elliptic genus and the dyon algebra of the $K3\times T^2$-compactified type II string theory. Notice that, as we discussed in the introduction, this does not necessarily mean all subgroups of $M_{24}$ have to be realised as an actual symmetry of the string theory at a certain point in the moduli space. But in the case when $g\in M_{24}$ does generate a symmetry group realised on some submanifold of the moduli space, we expect the above three properties to extend to it, once we drop the requirement that the action be fully geometric.
A construction of the new string theories obtained by orbifolding with these non-geometric symmetries, perhaps from the heterotic side by associating cycle shapes to heterotic lattices at enhanced symmetry points in a way similar to \cite{Chaudhuri1996}, will be extremely interesting. 
Hence, here we are in one of the interesting occasions when we know quite a lot about the answers (given by the above three conjectural formulas) before quite knowing what exactly the question is (what the new symmetries are that we have twisted the partition functions with).

Finally we finish with some discussions. 
First, a geometric interpretation of the above conjectures might be interesting in its own right. Recall that the $1/2$-BPS spectrum $1/\eta^{24}(\t)$ has an interpretation as  counting nodal curves in $K3$ \cite{Yau1996}, and the $1/4$-BPS index can be thought of as an appropriate counting of non-factorisable special Lagrangian cycles in $K3\times T^2$. The twisted objects give finer geometric information:  
While the $\eta$-products $1/\eta_g(\t)$ are now related to a certain counting of nodal curves in the corrresponding orbifolded $K3$, the twisted denominator formula should give a version of the ``twisted counting" of special Lagrangian cycles in $K3\times T^2$. 
Of course such a geometric interpretation of the partition functions is only applicable when $g$ generates a geometric symmetry. 

Second, following the long string derivation in \cite{Dijkgraaf1997} we get the following generating function for the twisted elliptic genera for the symmetric product of $K3$
$$
\sum_{N\geq 0} p^N {\cal Z}_g(S^N K3;\t,z) = \prod_{n>0, m\geq 0,\ell} \exp\left(\sum_{k=1}^\inf \frac{1}{k} c_{(g)^k}(4nm-\ell^2) \,(p^n q^m y^\ell)^k \right)\;.
$$
This differs from the twisted $1/4$-BPS partition function $1/\F_g$ by a factor
$$
\frac{y}{(1-y)^2}\prod_{m\geq1} \frac{(1-q^m)^4}{(1-q^m y)^2(1-q^m y^{-1})^2}\times \frac{1}{\eta_g(\t)}
$$
which also has a physical interpretation in terms  of the KK monopole and the center of mass degrees of freedom in the D1-D5-KK duality frame.

Third, as mentioned in the first section, by studying the twisted elliptic genera and relating them to the representations of $M_{24}$ we have learned that the representation should really be thought of as coming in conjugate pairs, rather than two duplicate copies as one might have concluded by just looking at the untwisted elliptic genus. A direct parallel of this in the dyon counting algebra is, instead of thinking of the spectrum as being generated by two copies of the same algebra and the generating function as being related to the denominator of the algebra  as $Z\sim (1/\text{den})^2$, it should really be thought of as $Z\sim  (1/\text{den})\times  (1/\overline{\text{den}})$. In particular, if we use the first interpretation, an interpretation of the twisted $1/4$-BPS indices as given by the denominator identity twisted by elements of $M_{24}$ (\ref{twisted_denominator}) of the original algebra will not be possible. 
Notice that this insight is not available, as far as the author can see, without making the connection between the ${\cal N}=4, d=4$ string theory with the sporadic group $M_{24}$.

Finally, as alert readers might have already noticed, apart from the twisted $1/4$-BPS indices of the un-orbifolded theory, we can also ask what the twisted and untwisted $1/4$-BPS indices of the theories orbifolded by the corresponding symmetries are. For the untwisted indices for the $\Z_{N}$-orbifolded theories, the partition function have been proposed in Ref.~\cite{JatkarJHEP0604:0182006,DavidJHEP0701:0162007} for $N=2,3,5,7$. Not surprisingly, these objects $1/\til \F_g(\O)$ are related to the  twisted partition functions $1/\F_g(\O)$ by an automorphic lift of the S-transformation $\t\to-\frac{1}{\t}$ from the genus one modular group $SL(2,\Z)$ into the genus two modular group $Sp(2,\Z)$. 
In Ref.~\cite{Cheng2008}, an interpretation in terms of a spectrum-generating function similar to the one reviewed in section \ref{The BPS Dyon Algebra and the Two Moonshines} was proposed for $N=2,3$ and in \cite{Govindarajan2009} for the case $N=4$. These algebras are not directly related to the original generalised Kac-Moody algebra (\ref{phi10})\footnote{For interested readers, they should be thought of as the analogues of the algebras ${\mathfrak{m}}_{g}, g \in {\bf M}$ constructed by S. Carnahan \cite{carnahan} as a part of the effort to prove Norton's conjecture on the modularity properties of the generalised McKay-Thompson series (\ref{generalisedMcKayT}) in the context of Monstrous Moonshine.}. As discussed above, from their relations with the sporadic group $M_{24}$ we now know that the dyon partition function should not be of the form $Z\sim (1/\text{den})^2$. We suspect that this previous prejudice might be related to the difficulty met in constructing the spectrum-generating algebra for $\Z_N$-orbifold with $N>4$. It might be worthwhile to re-examine this issue, now that we have much more information about the expected structures thanks to the insight the largest Mathieu group has brought into the structure of ${\cal N}=4$, $d=4$ string theories.


\section*{Acknowledgments}

I am deeply grateful to Clay Cordova, Atish Dabholkar, John Duncan, Jeff Harvey, Albrecht Klemm, Boris Pioline, Yuji Tachikawa, Cumrun Vafa, Erik Verlinde, and especially Ashoke Sen for extremely helpful discussions during the course of this work. I also thank {Eyjafjallaj\"okull} for making them possible. I would also like to thank LPTHE Jussieu, Universit\"at Bonn, and Universit\"at Heidelberg, where a part of this research was conducted, for their hospitality. 
This work is supported by the DOE grant number DE-FG02-91ER40654.

\appendix

\section{Character Tables} 
\label{Character Tables}

\medskip

 \begin{sidewaystable} \centering
 \resizebox{1.08\textwidth}{!}{
 \begin{tabular}[h!]{ccccccccccccccccccccccccccc}
 \toprule
cycle&$1^{24}$&$ 1^8 2^8 $&$ 1^6 3^6$&$1^4 5^4$&$1^4 2^2 4^4$&$ 1^3 7^3$&$1^3 7^3$&$ 1^2 2\, 4\, 8^2$& $1^2 2^2 3^2 6^2$&$1^2 \,11^2$&$1\;23$&$1\;23$&$1\;3\;5\;15$&$1\;3\;5\;15$&$1\;2\;7\;14$&$1\;2\;7\;14$&$12^2$&$6^4$&$4^6$&$3^8$&$2^{12}$&$2^2 10^2$&$3\,21$&$3\,21$&$2^4 4^4$&$2\,4\,6\,12$\\ \midrule
order&$1A$&$2A$&$3A$&$5A$&$4B$&$7A$&$\overline{7A}$&$8A$&$6A$&$11A$&$23A$&$\overline{23A}$&$15A$&$\overline{15A}$&
$14A$&$\overline{14A}$&$12B$&$6B$&$4C$&$3B$&$2B$&$10A$&$21A$&$\overline{21A}$&$4A$&$12A$\\ \midrule
&1&1&1&1&1&1&1&1&1&1&1&1&1&1&1&1&1&1&1&1&1&1&1&1&1&1\\
&23&7&5&3&3&2&2&1&1&1&0&0&0&0&0&0&-1&-1&-1&-1&-1&-1&-1&-1&-1&-1\\
&45&-3&0&0&1&$e_7$&$\bar e_7$&-1&0&1&-1&-1&0&0&$-e_7$&$-\bar e_7$&1&-1&1&3&5&0&$\bar e_7$&$e_7$&-3&0\\
&$\overline{45}$&-3&0&0&1&$\bar e_7$&$e_7$&-1&0&1&-1&-1&0&0&$-\bar e_7$&$-e_7$&1&-1&1&3&5&0&$e_7$&$\bar e_7$&-3&0\\
&231&7&-3&1&-1&0&0&-1&1&0&1&1&$e_{15}$&$\bar e_{15}$&0&0&0&0&3&0&-9&1&0&0&-1&-1\\
&$\overline{231}$&7&-3&1&-1&0&0&-1&1&0&1&1&$\bar e_{15}$&$e_{15}$&0&0&0&0&3&0&-9&1&0&0&-1&-1\\
&252&28&9&2&4&0&0&0&1&-1&-1&-1&-1&-1&0&0&0&0&0&0&12&2&0&0&4&1\\
&253&13&10&3&1&1&1&-1&-2&0&0&0&0&0&-1&-1&1&1&1&1&-11&-1&1&1&-3&0\\
&483&35&6&-2&3&0&0&-1&2&-1&0&0&1&1&0&0&0&0&3&0&3&-2&0&0&3&0\\
&770&-14&5&0&-2&0&0&0&1&0&$e_{23}$&$\bar e_{23}$&0&0&0&0&1&1&-2&-7&10&0&0&0&2&-1\\
&$\overline{770}$&-14&5&0&-2&0&0&0&1&0&$\bar e_{23}$&$e_{23}$&0&0&0&0&1&1&-2&-7&10&0&0&0&2&-1\\
&990&-18&0&0&2&$e_7$&$\bar e_7$&0&0&0&1&1&0&0&$e_7$&$\bar e_7$&1&-1&-2&3&-10&0&$\bar e_7$&$e_7$&6&0\\
&$\overline{990}$&-18&0&0&2&$\bar e_7$&$e_7$&0&0&0&1&1&0&0&$\bar e_7$&$e_7$&1&-1&-2&3&-10&0&$e_7$&$\bar e_7$&6&0\\
&1035&-21&0&0&3&$2 e_7$&$2 \bar e_7$&-1&0&1&0&0&0&0&0&0&-1&1&-1&-3&-5&0&$-\bar e_7$&$-e_7$&3&0\\
&$\overline{1035}$&-21&0&0&3&$2 \bar e_7$&$2 e_7$&-1&0&1&0&0&0&0&0&0&-1&1&-1&-3&-5&0&$-e_7$&$-\bar e_7$&3&0\\
&1035'&27&0&0&-1&-1&-1&1&0&1&0&0&0&0&-1&-1&0&2&3&6&35&0&-1&-1&3&0\\
&1265&49&5&0&1&-2&-2&1&1&0&0&0&0&0&0&0&0&0&-3&8&-15&0&1&1&-7&-1\\
&1771&-21&16&1&-5&0&0&-1&0&0&0&0&1&1&0&0
&-1&-1&-1&7&11&1&0&0&3&0\\
&2024&8&-1&-1&0&1&1&0&-1&0&0&0&-1&-1&1&1&0&0&0&8&24&-1&1&1&8&-1\\
&2277&21&0&-3&1&2&2&-1&0&0&0&0&0&0&0&0&0&2&-3&6&-19&1&-1&-1&-3&0\\
&3312&48&0&-3&0&1&1&0&0&1&0&0&0&0&-1&-1&0&-2&0&-6&16&1&1&1&0&0\\
&3520&64&10&0&0&-1&-1&0&-2&0&1&1&0&0&1&1&0&0&0&-8&0&0&-1&-1&0&0\\
&5313&49&-15&3&-3&0&0&-1&1&0&0&0&0&0&0&0&0&0&-3&0&9&-1&0&0&1&1\\
&5544&-56&9&-1&0&0&0&0&1&0&1&1&-1&-1&0&0&0&0&0&0&24&-1&0&0&-8&1\\
&5796&-28&-9&1&4&0&0&0&-1&-1&0&0&1&1&0&0&0&0&0&0&36&1&0&0&-4&-1\\
&10395&-21&0&0&-1&0&0&1&0&0&-1&-1&0&0&0&0&0&0&3&0&-45&0&0&0&3&0\\
 \bottomrule
  \end{tabular}}
  \caption{ \label{M24}\footnotesize{Character table of $M_{24}$. See \cite{atlas,james}. We adopt the naming system of \cite{atlas} and use the notation $e_n = \frac{1}{2}(-1+i\sqrt{n})$. }}
 \end{sidewaystable}

\newpage
\section{Modular Stuff}
\label{Modular Stuff}

{\it Theta Functions }
\vspace{.1cm}
\bea\notag
\th_1(\t,z) &=& -i q^{1/8} y^{1/2} \prod_{n=1}^\inf (1-q^n) (1-y q^n) (1-y^{-1} q^{n-1})\\ \notag
\th_2(\t,z) &=&  q^{1/8} y^{1/2} \prod_{n=1}^\inf (1-q^n) (1+y q^n) (1+y^{-1} q^{n-1})\\ \notag
\th_3(\t,z) &=&  \prod_{n=1}^\inf (1-q^n) (1+y \,q^{n-1/2}) (1+y^{-1} q^{n-1/2})\\
\th_4(\t,z) &=&  \prod_{n=1}^\inf (1-q^n) (1-y \,q^{n-1/2}) (1-y^{-1} q^{n-1/2})
\eea

\vspace{.2cm}\noindent
{\it Weak Jacobi Forms}
\vspace{.1cm}
\bea
\notag
\f_{0,1}(\t,z) &=& 4\,\left(\left(\frac{\th_2(\t,z)}{\th_2(\t,0)}\right)^2 +\left(\frac{\th_3(\t,z)}{\th_3(\t,0)}\right)^2 +\left(\frac{\th_4(\t,z)}{\th_4(\t,0)}\right)^2 \right)\\\notag
\f_{-2,1}(\t,z)&=& -\frac{\th_1^2(\t,z)}{\eta^6(\t)}
\eea

They are weak Jacobi forms of index one and weight $0$ and $-2$ respectively, and they generate the ring of weak Jacobi forms of even weight (Theorem 9.3 \cite{eichler_zagier}).

\vspace{.2cm}\noindent
{\it Higher Level}
\vspace{.1cm}

The family of congruence subgroups of the modular group $PSL(2,\Z)$ that are most relevant for this paper are 
\bea\notag
\G_0(N) &=& \bigg\{\bigg[\begin{array}{cc} a&b\\c&d\end{array}\bigg]\in PSL(2,\Z)  , c= 0 \text{ mod }N\;
\bigg\} \;.
\eea

A weight two modular form of $\G_0(N)$ is 
$$
\f_2^{(N)}(\t)=\frac{24}{N-1}\, q\pa_q\log\left(\frac{\eta(N\tau)}{\eta(\tau)}\right)=1+\frac{24}{N-1}\sum_{k>0}\s(k) (q^k -N q^{Nk}) \;,
$$
where $\s(k)$ is the divisor function $\s(k)=\sum_{d\lvert k}d$. 
This is the  combination of weight two, non-holomorphic Eisenstein series $G^*_2(\t)-NG^*_2(N\t)$, normalised such that the constant term in one. 

Apart from these so-called ``old-forms", for some $N$ there are also ``new-forms" which are weight two modular forms not of the above kind. Together they span the space of weight two modular forms of congruence subgroup $\G_0(N)$. 

For $N=11$, there is one such new form
\be\label{11newforms}
f_{11}(\t) = \eta^{2}(\t)\eta^{2}(11\t)\;.
\ee
A similar thing happens for $N=14,15$. And the respective new forms are 
\bea
f_{14}(\t) &=& \eta(\t)\eta(2\t)\eta(7\t)\eta(14\t)\\
f_{15}(\t) &=& \eta(\t)\eta(3\t)\eta(5\t)\eta(15\t)\;.
\eea

For $N=23$, there are two such new forms, one with coefficients in $\Z + \Z\frac{1-\sqrt{5}}{2}$ and the other obtained by replacing $\sqrt{5}$ by $-\sqrt{5}$
\bea\notag
\til f_{23,1}(\t)&=& q - \frac{1-\sqrt{5}}{2} q^2 -\sqrt{5} q^3 -  \frac{1+\sqrt{5}}{2} q^4 - (1-\sqrt{5}) q^5 - \frac{5-\sqrt{5}}{2} q^6 + \dotsi\\ \notag
\til f_{23,2}(\t)&=& q - \frac{1+\sqrt{5}}{2} q^2 +\sqrt{5} q^3 -  \frac{1-\sqrt{5}}{2} q^4 - (1+\sqrt{5}) q^5 - \frac{5+\sqrt{5}}{2} q^6 + \dotsi\;.
\eea

Here we use the basis such that they are Hecke Eigenforms. For the actual computation it is more convenient to use the basis
\be\label{23newforms}
f_{23,1}(\t) = \til f_{23,1}(\t)+\til f_{23,2}(\t)\quad,\quad f_{23,2}(\t) = \frac{1}{\sqrt{5}}\left(\til f_{23,1}(\t)-\til f_{23,2}(\t)\right)\;.
\ee

See \cite{modi} and Chapter 4.D. of \cite{from_number} for more details on these modular forms.

A discussion about the ring of weak Jacobi forms of higher level can be found in Ref.~\cite{aoki}. From its Proposition 6.1, we conclude that weight zero, index one weak Jacobi forms of congruence subgroup $\G_0(N)$ are linear combinations of $\f_{0,1}(\t,z)$ and  $\f_{-2,1}(\t,z)\times f_N(\t)$, where $f_N(\t)$ is a weight two modular form of $\G_0(N)$ and is hence linear combinations of the above ``old" and ``new" forms. 

\section{Formulas for Twisted Elliptic Genera}
\label{Formulas for Twisted Elliptic Genera}

In this appendix we collect examples of elliptic genera of the $K3$ CFT twisted by elements of $M_{24}$. Unlike the examples discussed in Section \ref{Twisted Elliptic Genera}, in these examples the elements $g\in M_{24}$ do not 
have known interpretation as acting on the classical geometry of $K3$ surfaces. 
\bea\notag
{\cal Z}_{11A}(\t,z) &=& \frac{1}{6}\f_{0,1}(\t,z) +\f_{-2,1}(\t,z) \left( \frac{11}{6}\f_2^{(11)}(\t) -\frac{22}{5} f_{11}(\t)\right) \\ \notag
{\cal Z}_{23A}(\t,z) &=& \frac{1}{12}\f_{0,1}(\t,z) +\f_{-2,1}(\t,z) \left( \frac{23}{12}\f_2^{(23)}(\t) -\frac{23}{22} f_{23,1}(\t) -\frac{161}{22} f_{23,2}(\t)\right) \\ \notag
{\cal Z}_{14A}(\t,z) &=& \frac{1}{12}\f_{0,1}(\t,z) +\f_{-2,1}(\t,z) \left(- \frac{1}{36}\f_2^{(2)}(\t) 
-\frac{7}{12}\f_2^{(7)}(\t)+\frac{91}{36}\f_2^{(14)}(\t)
-\frac{14}{3} f_{14}(\t)\right) \\ \notag
{\cal Z}_{15A}(\t,z) &=& \frac{1}{12}\f_{0,1}(\t,z) +\f_{-2,1}(\t,z) \left(- \frac{1}{16}\f_2^{(3)}(\t) 
-\frac{5}{24}\f_2^{(5)}(\t)+\frac{35}{16}\f_2^{(15)}(\t)
-\frac{15}{4} f_{15}(\t)\right) \\\notag
{\cal Z}_{2B}(\t,z) &=& \f_{-2,1}(\t,z)\left(-2\f_2^{(2)}(\t)+4\f_2^{(4)}(\t)\right)\\ \notag
&=&16\, \f_{-2,1}(\t,z)\,
q\pa_q \log\left(\frac{\eta(\t)\,\eta^2(4\t)}{\eta^3(2\t)}\right)\\ \notag
{\cal Z}_{4A}(\t,z) &=& \f_{-2,1}(\t,z)\left(\frac{1}{3}\f_2^{(2)}(\t)-3\f_2^{(4)}(\t)+\frac{14}{3}\f_2^{(8)}(\t)\right)\\
&=&8\, \f_{-2,1}(\t,z)\,
q\pa_q \log\left(\frac{\eta(2\t)\,\eta^2(8\t)}{\eta^3(4\t)}\right)
\eea
The weight two modular forms in the above formulas are discussed in the last Appendix. 
As checks of these formulas, one can easily verify that all the Fourier coefficients are integers. 

One curious observation is, one can show that the McKay-Thompson series of the form in (\ref{guess1}) is a weak Jacobi form of the group $\,\G_0(\text{ord}(g))$ if and only the element $g$ is not only an element of $M_{24}$ but also an element of $M_{23}$. This fact supports the possibility that for all conjugacy class of $M_{23}$, there is a point in the moduli space where the symmetry is realised as a symmetry of the CFT. 
For the two examples ($g=2B, 4A$) of elements that are in $M_{24}$ but not in $M_{23}$, slightly more subtle modular properties of the corresponding twisted elliptic genus ${\cal Z}_{g}(\t,z)$ are discussed in \cite{Gaberdiel2010}\footnote{Paragraph rewritten in the second version. We thank M. Gaberdiel for correspondence on this point.}.

\bibliography{mathieu.bib}{}

\end{document}